\documentclass[aps, 12pt,superscriptaddress, amsmath, amssymb, noshowpacs, preprint,preprintnumbers]{revtex4}
\usepackage{amsmath}
\usepackage{amsfonts}
\usepackage{amssymb}
\usepackage{textcomp}
\usepackage{graphicx}
\usepackage{bm}
\usepackage{color}

\usepackage{epsfig}
\usepackage{amsfonts,amsmath,amssymb}

\def\beq{\begin{equation}}
\def\eeq{\end{equation}}
\def\beqn{\begin{eqnarray}}
\def\eeqn{\end{eqnarray}}
\newcommand{\be}{\begin{equation}}
\newcommand{\ee}{\end{equation}}
\newcommand{\ba}{\begin{eqnarray}}
\newcommand{\ea}{\end{eqnarray}}
\newcommand{\ben}{\begin{enumerate}}
\newcommand{\een}{\end{enumerate}}

\newcommand{\tz}{\tilde{z}}
\newcommand{\ts}{\tilde{\sigma}}

\newcommand{\p}{\partial}
\newcommand{\ph}{\varphi}

\newcommand{\la}{\langle}
\newcommand{\ra}{\rangle}

\newcommand{\rar}{\rightarrow}
\newcommand{\eps}{\epsilon}
\begin{document}

\title{Baryon as dyonic instanton-II.  \\ Baryon mass versus chiral condensate}

\preprint{ITEP-TH 26/13}

\author{A. Gorsky}
\affiliation{Institute for Theoretical and Experimental Physics (ITEP), \\ Moscow, Russia
}
\affiliation{Moscow Institute of Physics and Technology (MIPT),\\
Dolgoprudny, Russia}
\author{P. N. Kopnin}
\affiliation{Institute for Theoretical and Experimental Physics (ITEP), \\ Moscow, Russia
}
\author{A. Krikun}

\affiliation{Institute for Theoretical and Experimental Physics (ITEP), \\ Moscow, Russia
}

\affiliation{All-Russia research institute of automatics (VNIIA), \\
Center for fundamental and applied research (CFAR),  \\
 Moscow, Russia \\
}

\begin{abstract}
We discuss the description of baryon as the dyonic instanton in holographic QCD. The solution generalizes
the Skyrmion taking into account the infinite tower of  vector and axial mesons as well as the chiral
condensate. We construct the solution with unit baryon charge and study the dependence of its mass on the chiral condensate. The elegant explanation of the Ioffe's  formula has been found and we speculate on the relation between physical scales of the chiral and conformal symmetry breaking.
\end{abstract}

\maketitle

\section{Introduction}

In \cite{gk} two of us have introduced the generalization of the Skyrmion model for the baryon. The
idea was to take into account the infinite tower of the vector and axial mesons and the chiral condensate.
These data are most effectively packed into the 5d holographic model for QCD \cite{hard-wall} in the curved geometry.  The expansion of
5d gauge fields with $SU(N_f)_L\times SU(N_f)_R$ gauge group  into KK modes yields the whole tower of the massive mesons while the chiral condensate is encoded in the boundary condition for the bifundamental scalar.  In fact the holographic QCD is just the Chiral Lagrangian
supplemented by the tower of the massive mesons in a particular way. The conventional Skyrmion is just the solitonic configuration built from the pions only however there were some attempts to include the vector mesons into the solution (see \cite{mei} for review). In particular the
stabilization of the Skyrmion via the $\omega$ - mesons has been suggested \cite{nappi} which substitutes the Skyrme term mechanism. More recently the Skyrmion has been discussed in 5d model in the flat space and it was found that
the vector and axial mesons strongly influence the Skyrmion solution \cite{sutc}.
In \cite{gk} the generalization of the Skyrmion solution with highly nontrivial role of the whole tower of the axial and vector mesons  has been found. Some aspects of the influence of the  meson tower on the Skyrmion solution have been discussed in \cite{rho}.

The aim of this study is to treat the longstanding puzzle concerning the role of the QCD chiral condensate in the baryon mass generation from a new perspective using the approach developed in \cite{gk}. This subject was triggered by
the sum rule calculation of the baryon mass  \cite{Ioffe}  resulted in the following surprising formula
\beq
\label{Form_Ioffe}
M^3 = -8 \pi^2 \la \bar{q}q \ra
\eeq
which is valid with reasonable accuracy. It implies that the huge amount of the baryon mass is due to the chiral condensate. However
during the 30 years there was no any convincing analytical calculation which would put this formula at the firm ground. Moreover the lattice calculations at the non-vanishing temperature and density added the controversy to the subject. There are lattice calculation near the phase transition temperature which shows no substantial change of the baryon mass at the transition point (\cite{glozman} and refs therein). The calculation at the non-vanishing density indicates some dependence of the baryon mass at the scale where the transition to the
half-Skyrmion phase happens \cite{Ma} however there is no strong dependence at the restoration of the chiral symmetry point. It was suggested that the naive formula
\beq
\label{linear}
M_B= m_0 + \Delta( \la \bar{q}q \ra),
\eeq
where $m_0$ is condensate independent, well interpolates the physics. The similar formula has been
also used in the parity-doublet model and the Chiral Quark-Soliton model (see \cite{diak} for a review). In the
latter it is assumed that the baryon consists of the three constituent quarks whose masses are related to the
chiral condensate. The constituent quarks are in a topologically nontrivial pion field which provides their localization as a bound state. The total baryon mass is interpreted as the sum of  constituent masses
and contribution due to  the shift of the negative continuum induced by the nontrivial pion field.

The conventional  Skyrmion was identified as the instanton solution at the 4d slice of the 5d AdS-like space \cite{hard-wall} where the holographic RG coordinate plays the role of the Euclidean time direction. It can be also considered along the Atiah-Manton picture  \cite{Atiah-Manton} as the
instanton trapped by the domain wall localized in the holographic coordinate. The holographic perspective suggests the proper generalization of the instanton solution which would take into account the massive mesons and chiral condensate. It is similar to the dyonic instanton solution found in \cite{lt} in the 5d SUSY gauge theory. Its key feature is  fixing of the instanton size by the non-vanishing second charge while the mass scale is provided by the vev of the adjoint scalar.  The generalization of the Lambert-Tong solution appropriate to our purpose is not immediate and deserves for some care \cite{gk}. Since we are working with the gauge theory at the flavor probe branes the
gauge group is the product $SU(N_f)_L\times SU(N_f)_R$  and the scalar lives in the bifundamental representation of the flavor gauge group instead of adjoint as in the dyonic instanton for the color group. Moreover we have to take into account the nontrivial metric in 5d space which makes the analytical solution to the equation of motion quite difficult. In \cite{gk} it was shown that the solution of the dyonic instanton type exists at the particular choice of the boundary conditions. It has strong similarity with the Atiyah-Manton picture of the domain wall localized at the holographic coordinate.

In this paper we will do the next step within this approach and investigate the dependence of the baryon mass on the chiral condensate in the generalized Skyrmion  suggested in \cite{gk}. To this aim we consider more general boundary conditions which provide more freedom in the solution. The chiral condensate was introduced via the boundary condition for the bifundamental scalar while we assume no explicit quark mass term. We shall find the numerical solution to the equations of motion and demonstrate analytically that it carries one unit of the baryon charge. Its stabilization is dictated by the chiral condensate and the confinement scale. Remarkably our dyonic instanton solution yields the elegant solution to the longstanding puzzle concerning the origin of the  Ioffe's formula  (\ref{Form_Ioffe}). It turns out that the dependence of the baryon mass on the chiral condensate has two branches. At small values of the condensate the mass is condensate independent  while starting at some critical value  it 
grows according to the Ioffe's formula.
The analysis shows that the nature somehow dynamically selects
the value of the condensate just at the intersection of two branches. Therefore we obtain the natural
explanation of the Ioffe's formula while at small values of the condensate
the baryon mass becomes almost condensate independent.

The paper is organized as follows. In Section \ref{2} we implement the above ideas in the simple ``hard-wall'' AdS/QCD model, find nontrivial solution with unit baryon charge and study the dependence of the baryon mass on the value of the chiral condensate. Section \ref{3} is devoted to the comments on the  brane picture for the  solution obtained. We conclude with discussion in Section \ref{4} and give the details of numerical calculation in Appendix.

\section{Hard wall holographic model \label{2}}
\subsection{The model}
In order to realize the ideas mentioned in the Introduction quantitatively we use the simple holographic AdS/QCD model with ``hard wall'' \cite{hard-wall, pomarol}. The bulk space is five-dimensional and has the pure $AdS^5$ metric.
\begin{equation}
\label{metric}
ds^2 = \frac{1}{z^2} (dt^2 - dz^2 - dx_i^2), \quad i=1 \dots 3, \quad z<z_m
\end{equation}
The conformal symmetry is broken by the hard hard wall placed at $z=z_m$. Here and further in the paper we rescale the curvature radius of the space to 1.

The field content of the model includes two gauge fields $L_\mu^a$ and $R_\mu^a$ in the adjoint representation of the flavor group $U(N_f)$ dual to the left and right quark currents, respectively. Their vector and axial linear combinations will describe the full towers of vector and axial mesons in the model. Another field is a bifundamental scalar $X^{\alpha \beta}$ which is dual to the quark bilinear operator $\bar{q}^\alpha q^\beta$, whose vacuum expectation value is a chiral condensate. In what follows we take $N_f = 2$ because this leads to relatively simple expressions and should be enough to describe the main qualitative features of the baryon (nucleon) physics. The action reads as (we use the notation of \cite{Krik1,Krik2,tensor})
\begin{equation}
\label{action}
S = \int d^3 x dt dz \left\{ \frac{1}{z} \left( - \frac{1}{4 g_5^2} \right) (F_L^2 + F_R^2) + \frac{\Lambda^2}{z^3} (D X)^2 + \frac{\Lambda^2}{z^5} \  3|X|^2 \right\},
\end{equation}
where the covariant derivative is $D_\mu X = \p_\mu X - i L_\mu X + i X R_\mu $. The constants in the action can be fixed by matching the two-point functions calculated in the holographic model to the leading terms in the sum rule approach \cite{hard-wall, Krik2}: $\Lambda^2 = \frac{3}{g_5^2}$, $\frac{1}{g_5^2} = \frac{N_c}{12 \pi^2}$. Apart of the Yang-Mills part of the action one needs to consider the Chern-Simons part, which is
\begin{equation}
\label{CS}
S_{CS} = \frac{N_c}{24 \pi^2} \int \frac{3}{2} \left \{\hat{L} \ tr (F_L \tilde{F_L}) - \hat{R}  \ tr ( F_L \tilde{F_L}) \right \}.
\end{equation}
Here we mark with a hat the Abelian (related to the Cartan subalgebra of $SU(N_f)$) parts of the gauge fields. One should recall \cite{Sakai-baryon, Wulzer} that as the Abelian part of the gauge field is dual to the $U(1)$ quark current, the vector combination $\hat{L}_\mu + \hat{R}_\mu = \hat{V}_\mu$ is dual to the baryon number current. Now we can see from (\ref{CS}) that the source for the temporal component of the baryon current -- the baryon charge -- in our model is
\begin{equation}
\label{baryon}
Q_B =\frac{1}{16 \pi^2} \int d^3 x dz \epsilon^{\mu \nu \lambda \rho} \left[F_L^{\mu \nu} F_L^{\lambda \rho} - F_R^{\mu \nu} F_R^{\lambda \rho} \right].
\end{equation}
This is identified with the difference between four-dimensional topological charges of the left and right nonabelian field configuration in the constant time slice. Thus in order to describe the baryon we need to find the solution to the equations of motion of the model which has nontrivial topological charge -- the instanton. This task turns out to be nontrivial even in the simple metric like (\ref{metric}), since the usual way of obtaining instanton solution via the Bogomolny equations fails in the curved background. The thorough study of such solutions in Sakai-Sugimoto model was performed recently in \cite{Bolognesi}.

\subsection{The ansatz}
In order to construct the three-dimensional spherical symmetric solution we use the ansatz similar to the Witten cylindric ansatz \cite{Witten_inst} with the holographic coordinate $z$ along the axis of the cylinder and the 3D radius $r$ in the slice of constant $z$. Now the model can be rewritten in terms of two-dimensional fields:
\begin{align}
\label{ansatz}
L_j^a &= -\frac{1 + \xi_2 (r,z) + \eta_2 (r,z)}{r} \epsilon_{jak} \frac{x_k}{r} + \frac{\xi_1 (r,z) + \eta_1 (r,z)}{r} \left( \delta_{ja} - \frac{x_j x_a}{r^2} \right) \\
\notag
&  + \big(V_r(r,z) + A_r(r,z)\big)  \frac{x_j x_a}{r^2}, \\
\notag
L_5^a &= \big(V_z(r,z) + A_z(r,z)\big) \frac{x_a}{r}, \\
\notag
R_j^a &= -\frac{1 - \xi_2 (r,z) + \eta_2 (r,z)}{r} \epsilon_{jak} \frac{x_k}{r} + \frac{\xi_1 (r,z) - \eta_1 (r,z)}{r} \left( \delta_{ja} - \frac{x_j x_a}{r^2} \right) \\
\notag
&  + \big(V_r(r,z) - A_r(r,z) \big) \frac{x_j x_a}{r^2}, \\
\notag
R_5^a &= \big(V_z(r,z) - A_z(r,z)\big) \frac{x_a}{r},\\
X &= \chi_1(r,z) \frac{\mathbf{1}}{2} + i \chi_2(r,z) \frac{\tau^a x^a}{r},
\end{align}
where $\tau^a$ are the generators of $SU(2)$ obeying the commutation relation $[\tau^a \tau^b] = - i \epsilon^{abc} \tau^c$. In this notation the covariant derivative $D_\mu X$ takes the convenient form
\begin{align}
D_i X = \frac{\mathbf{1}}{2} \frac{x_i}{r} & \left[ \chi_1' + \chi_2 A_r \right] - i \tau^a \left(\delta_{ai} - \frac{x_i x_a}{r^2} \right)  \frac{1}{r} \left[\chi_2 \eta_2 + \chi_1 \eta_1 \right]  \\
\notag
+ i \tau^a \ \frac{x_i x_a}{r^2} \ & \left[ \chi_2' - \chi_1 A_r \right] + i \tau^a \ \eps_{iak} \frac{x_k}{r}  \ \frac{1}{r} \left[ \chi_2 \xi_1 - \chi_1 \xi_2 \right], \\
\notag
D_5 X = \frac{\mathbf{1}}{2} & \left[ \dot{\chi}_1 + \chi_2 A_z \right]  + i \tau^a \frac{x_a}{r}  \left[ \dot{\chi}_2 - \chi_1 A_z \right].
\end{align}
We see that the scalar interacts only with the axial combination of the gauge fields and the ansatz fixes the gauge up to the $U(1)$ subgroup. The remaining $U(1)$ gauge symmetry can be seen in the two-dimensional covariant derivative terms. Indeed, the complex combination $\chi_1 + i \chi_2$ behaves as a charged scalar with respect to the Abelian $A_\alpha$ (this was observed earlier in \cite{Wulzer4}). The full action of the model in the ansatz (\ref{ansatz}) is
\begin{align}
\label{action-scalars}
S = \frac{N_c}{6 \pi} \int d r dt dz \bigg\{ - &\frac{r^2}{z} \frac{1}{2} (\p_z A_r - \p_r A_z)^2 - \frac{r^2}{z} \frac{1}{2} (\p_z V_r - \p_r V_z)^2 \\
\label{kinetic-scalars}
-  \frac{1}{z}  \Big[ &(\eta_2' - V_1 \xi_1 - A_1 \eta_1 )^2  + (\xi_2' - V_1 \eta_1 - A_1 \xi_1 )^2 \\
\notag
+ & (\xi_1' + V_1 \eta_2 + A_1 \xi_2 )^2 + (\eta_1' + V_1 \xi_2 + A_1 \eta_2 )^2 \\
\notag
+ & (\dot{\eta}_2 - V_2 \xi_1 - A_2 \eta_1 )^2 + (\dot{\xi}_2 - V_2 \eta_1 - A_2 \xi_1 )^2 \\
\notag
+ & (\dot{\xi}_1 + V_2 \eta_2 + A_2 \xi_2 )^2 + (\dot{\eta}_1 + V_2 \xi_2 + A_2 \eta_2 )^2 \Big] \\
\label{gauge-potentials}
- & \frac{1}{z r^2} \left(1 - \xi_1^2 - \eta_2^2 - \eta_1^2 - \xi_2^2 \right)^2 - \frac{4}{z r^2} \left(\xi_1 \eta_1 + \xi_2 \eta_2 \right)^2 \\
\label{kinetic-chi}
- \frac{3 r^2}{z^3} &\Big[ \big(\chi'_1 + \chi_2 A_1\big)^2  + \big(\dot{\chi}_1 + \chi_2 A_2\big)^2 \\
\notag
+ & \big(\chi'_2 - \chi_1 A_1)^2  + \big(\dot{\chi}_2 - \chi_1 A_2\big)^2 \Big]\\
\label{scalar-potentials}
- \frac{6}{z^3} & \Big[ (\chi_1 \eta_1 + \chi_2 \eta_2)^2 + (\chi_1 \xi_2 - \chi_2 \xi_1)^2   \Big] \\
+ \frac{9 r^2}{z^5} & (\chi_1^2 + \chi_2^2).
\end{align}

Our goal is to find a configuration of nonabelian gauge and scalar fields which has a finite energy in the four-dimensional $t=const$ slice. This means that on the boundaries of the integration region the kinetic terms (\ref{action-scalars}),(\ref{kinetic-scalars}),(\ref{kinetic-chi}) as well as the potential terms (\ref{gauge-potentials}),(\ref{scalar-potentials}) should vanish fast enough. To proceed with the analysis of the problem it is convenient to introduce the phases and moduli of the scalars as variables
\begin{align}
\label{phases}
\eta_1 &= \phi \cos(\theta) \cos(\alpha), & \eta_2 &= \phi \cos(\theta) \sin(\alpha), \\
\notag
\xi_1 &= \phi \sin(\theta) \cos(\beta), & \xi_2 &= \phi \sin(\theta) \sin(\beta), \\
\notag
\chi_1 &= \chi \cos(\gamma), & \chi_2 &= \chi \sin(\gamma).
\end{align}

Rewriting the expression (\ref{action-scalars}) in terms of these phases and denoting $\alpha - \beta = \omega$, we get the convenient expression for the energy of the solution
\begin{align}
\label{energy-phases}
E =  \frac{N_c}{6 \pi} \int dr \ dz \ \Bigg\{  & \frac{2}{z} (\p_z \phi)^2 + \frac{2}{z} (\p_r \phi)^2\\
\notag
&+ \frac{1}{z} \phi^2 \Big[ \cos(\theta)^2 (A_z - \p_z \alpha)^2 + \sin(\theta)^2 (A_z - \p_z \alpha + \p_z \omega)^2 + \cos(\omega)^2 (\p_z \theta)^2   \\
\notag
& \qquad  + ( \sin(\omega) \p_z \theta + V_z)^2 + (2 A_z - 2 \p_z \alpha + \p_z \omega) \sin(2\theta) V_z \cos(\omega) \Big] \\
\notag
&+ \frac{1}{z} \phi^2 \Big[ \cos(\theta)^2 (A_r  - \p_r \alpha)^2 + \sin(\theta)^2 (A_r  - \p_r \alpha + \p_r \omega)^2 + \cos(\omega)^2 (\p_r \theta)^2   \\
\notag
& \qquad  + ( \sin(\omega) \p_r \theta + V_r)^2 + (2 A_r - 2 \p_r \alpha + \p_r \omega) \sin(2\theta) V_r \cos(\omega) \Big] \\
\notag
+& \frac{r^2}{2 z} (\p_z A_r - \p_r A_z)^2 + \frac{r^2}{2 z} (\p_z V_r - \p_r V_z)^2  \\
\notag
+& \frac{1}{r^2 z} (1 - \phi^2)^2 + \frac{1}{r^2 z} \phi^4 \sin(2 \theta)^2 \cos(\omega)^2 \\
\notag
+ & \frac{3 r^2}{z^3} (\p_z \chi)^2 + \frac{3 r^2}{z^3} \chi^2 (\p_z \gamma - A_z)^2  \\
\notag
+ & \frac{3 r^2}{z^3} (\p_r \chi)^2 + \frac{3 r^2}{z^3} \chi^2 (\p_r \gamma - A_r)^2   \\
\notag
+ & \frac{6}{z^3} \chi^2 \phi^2 \ \Big[ \cos(\theta)^2 \cos(\gamma -\alpha)^2 + \sin(\theta)^2 \sin(\gamma - \alpha + \omega)^2 \Big]  \\
\notag
- & \frac{9 r^2}{ z^5} \chi^2 \Bigg\}
\end{align}

For (\ref{gauge-potentials}) one has
\begin{equation}
\label{potential1}
 \frac{1}{z r^2}  \Big[(\phi^2 - 1)^2 + \phi^4 \sin(2 \theta)^2 \cos(\omega)^2 \Big].
\end{equation}
 Since this potential is positively definite, the lowest energy solution is defined by the two terms separately. The first leads to $\phi = 1$, the second -- to either $\theta = \frac{\pi}{2} n$ or $\omega = \frac{\pi}{2} + \pi m$ ($m,n \in \mathbb{Z}$). Next, the potential (\ref{scalar-potentials}) turns to
\begin{align}
\label{potential2}
\frac{6}{z^3} \chi^2 \ph^2 \big[ \cos(\theta)^2 \ \cos(\gamma - \alpha)^2 + \sin(\theta)^2 \ \sin(\gamma  - \alpha + \omega))^2 \big].
\end{align}
and consequently in the vacuum state either $\sin(2 \theta) = 0$ and
\begin{align*}
\gamma - \alpha = \frac{\pi}{2} + \pi n, & \quad n \in \mathbb{Z} &  \mbox{ for } &\cos(\theta) = 0, \\
\gamma-\alpha = -\omega   + \pi n, & \quad n \in \mathbb{Z} & \mbox{ for } &\sin(\theta) = 0,
\end{align*}
or  $\sin(2 \theta) \neq 0$ and
\begin{align*}
\omega = \frac{\pi}{2} + \pi n, \qquad n \in \mathbb{Z}.
\end{align*}

The solution with minimal energy should have the vacuum asymptotics with vanishing potentials (\ref{potential1}) and (\ref{potential2}) at each boundary of the integration region: $z=0$, $z=z_m$, $r=0$ and $r \rar \infty$. Moreover, to find a solution with nonzero topological charge we set the different vacuum asymptotics at different boundaries. This allows for the solution, which describes the tunneling between different vacuum states, namely the instanton. Apart of that one should check that the energy (\ref{energy-phases}) is finite at the boundaries. Before proceeding with the definition of the boundary values of phases we note, that in (\ref{energy-phases}) the phase $\alpha$ is subject to the gauge symmetry, so we can fix the gauge by taking the constant value for it:
\begin{equation}
\label{alpha}
\alpha = -\frac{\pi}{2}
\end{equation}

Studying the equations of motion on the boundaries we find, that the equation for the modulus $\chi$ is always satisfied by
\begin{equation}
\label{chi}
\chi = m z + \sigma z^3.
\end{equation}
This is not surprising at all because the field $X$, whose modulus is described by $\chi$, is holographically dual to the quark bilinear operator $(\bar{q}q)$ \cite{hard-wall, gubser-klebanov}. The form of the solution is dictated by the canonical dimension $\Delta$ of this operator: one branch behaves as $z^{4-\Delta}$ and the other as $z^{\Delta}$. The two dimensional coefficients are related to the source of the operator (which is for  $(\bar{q}q)$ the quark mass $m$) and the vacuum expectation value $\la \bar{q}q \ra$, namely the quark condensate. The parameter $\sigma$ is related to the quark condensate as \cite{Krik1}
\begin{equation}
\label{sigma}
\sigma = \frac{N_f}{3 \Lambda^2}  \langle \bar{q}q \rangle \approx (460 Mev)^3
\end{equation}
In what follows we study the chiral limit $m = 0$. We anticipate that the properties of the baryons are not changed significantly  in this limit.

 At the boundary $z=0$ we choose the vacuum state with $\phi=1$, $\theta = 0$. The absence of the sources for the vector currents in the system under consideration leads on the holographic side to the definite asymptotic of the vector fields $A_\alpha, V_\alpha \sim z^2$ \cite{hard-wall, gubser-klebanov}. From (\ref{energy-phases}) we find that the vanishing of the energy at $z=0$ allows  the finite derivative of $\omega$ along the boundary, so we consider the change of $\omega$ from $-\frac{\pi}{2}$ to $\frac{\pi}{2}$ on the way from $r=0$ to $r \rar \infty$.

The boundary $r=0$ is therefore characterized by $\omega = - \frac{\pi}{2}$ and still $\phi=1$. The finiteness of the energy requires that $A_r, V_r, A_z =0$, but the derivative of $\theta$ can be finite provided that $V_z = \p_z \theta$. Consequently we can consider the change of $\theta$ from $0$ to $\frac{\pi}{2}$ along the $r=0$ boundary while $z$ runs from $0$ to $z_m$.

The similar picture is observed on the boundary $r \rar \infty$. Here we take $\omega = \frac{\pi}{2}$ and $\phi=1$. Although the energy is finite only for $A_r, V_r, A_z =0$, the derivative of $\theta$ is related to $V_z$ as $V_z = -\p_z \theta$ being finite. Again we can consider the change of $\theta$ from $0$ to $\frac{\pi}{2}$ along the $r \rar \infty$ boundary.

On the remaining boundary $z=z_m$ we impose the conditions which force the Lagrangian to vanish. Although this is not necessary for the finiteness of the energy, in more complicated models, for instance for the soft-wall model \cite{soft-wall}, these boundary conditions would allow to extend the region of integration on $z$ to infinity while keeping the energy finite. Therefore the chosen boundary conditions on the IR wall ensure us that our treatment can be generalized to other holographic QCD models without qualitative changes. Thus on the boundary $z=z_m$ we take $\theta = \frac{\pi}{2}$, $V_z, V_r, A_z = 0$ and $A_r = - \p_r \omega$. Similarly to the $z=0$ boundary, the phase $\omega$ changes from  $-\frac{\pi}{2}$ to $\frac{\pi}{2}$.  The potential (\ref{potential2}) requires then, that $\gamma$ changes from $\pi$ to $0$ (taking (\ref{alpha}) into account).

\begin{table}[h!]
\begin{tabular}{|c|c|c|c|c|}
 \hline
$\quad r \quad$ & $0 \rar \infty$ & $\infty$ & $\infty \rar 0 $ & $ 0 $\\ \hline
$z $ & $ 0 $ & $ 0 \rar z_m $ & $ z_m $ & $ z_m \rar 0 $\\ \hline \hline
$\theta $ & $ 0 $ & $ 0 \rar \frac{\pi}{2} $ & $ \frac{\pi}{2} $ & $ \frac{\pi}{2} \rar 0$ \\ \hline
$\omega $ & $ -\frac{\pi}{2} \rar \frac{\pi}{2} $ & $ \frac{\pi}{2} $ & $ \frac{\pi}{2} \rar -\frac{\pi}{2} $ & $ -\frac{\pi}{2} $\\ \hline
$\gamma $ & $ \pi \rar 0 $ & $ 0 $ & $ 0 \rar \pi $ & $ \pi$ \\ \hline
\end{tabular}
\caption{The boundary values for the phases (\ref{phases}), which describe the solution with unit baryon charge.}
\end{table}

 At the end of the day we have constructed the self consistent boundary values of the phases which correspond to the different vacuum asymptotics of the solution at different boundaries. We summarize these conditions in the Table 1. As the solution, which corresponds to these boundary values, interpolates between different vacuum states we anticipate that it should have nonzero topological number similarly to the flat space instanton. Indeed, the baryon charge (\ref{baryon}) in terms of the phases (\ref{phases}) can be expressed in the form
\begin{align*}
Q_B =\frac{1}{\pi} \bigg \{  \int \limits_{0}^{z_m} dz \Big[& (\ph^2-1)
\left(A_z - \p_z \alpha \right) + \ph^2 \left(V_z \sin(2\theta) \cos(\omega) - \sin(\theta)^2 \p_z \omega \right)  \Big]_{r=0}^{r \rar \infty}  \\
- \int \limits_{0}^{\infty} dr \Big[& (\ph^2-1) \left(A_r - \p_r \alpha \right) + \ph^2 \left(V_r \sin(2\theta) \cos(\omega) - \sin(\theta)^2 \p_r \omega \right)  \Big]_{z=0}^{z=z_m} \bigg \}.
\end{align*}
With the boundary values from Table 1 this expression reduces to
\begin{align}
\label{charge1}
Q_B =\frac{1}{\pi} (\omega|_{z=z_m, r\rar \infty} - \omega|_{z=z_m, r=0 }) = 1
\end{align}
and we see that the solution under consideration is indeed a baryon. We should stress here that the fact that the topological (baryon) charge is integer is by no means trivial in $AdS$ space. In \cite{Maldacena} it was pointed out that the usual considerations leading to the result that the topological number is integer are spoiled in $AdS$, because having a boundary this space has a topology of the disk. Moreover the solutions with half-integer topological number, merons, are easier to obtain, then the integer ones \cite{merons}. On the other hand, given that the baryon charge under consideration is actually not the topological number, but the difference of topological numbers of left and right fields, one can show \cite{WulzerTopCharge} that the baryon charge is integer if the special boundary conditions on the gauge fields are satisfied on the IR boundary. In our treatment the integer baryon charge is the consequence of the discrete number of vacuum states of the theory and no additional constraints is 
needed to ensure that. Moreover, as we already pointed out earlier, our treatment allows the generalization to the other geometries, even extended to $z \rar \infty$, and the integer baryon number of the solution under consideration is not affected by this generalization at all.

In this section we used only one possible configuration of the boundary values for phases. Let us comment on the other possibilities. First of all, let us note, that we would not be satisfied with only one phase, $\omega$ or $\theta$, changing and the other constant. If $\omega$ was constant the baryon charge (\ref{charge1}) would be obviously zero and if $\theta$ was constant then the contribution from the $z=0$ boundary would cancel the contribution from $z=z_m$ boundary. Thus the change of both phases should be considered. From the other hand, we could take $\theta=0$ at $r=0$, $\omega = \frac{\pi}{2}$ at $z=0$, $\omega = -\frac{\pi}{2}$ at $z=z_m$, and $\theta= \frac{\pi}{2}$ at $r \rar \infty$. Although one can check, that in this arrangement the baryon charge would be equal to unity, this phase dynamics requires finite derivative of $\theta$ along the $z=0$ boundary. With finite energy (\ref{energy-phases}) this can be achieved only if $\p_r \theta|_{z=0} =- V_r|_{z=0}$, but the absence of sources for 
the vector current tells us that $V_r|_{z=0} \sim z^2$. Hence $\p_r \theta|_{z=0}=0$ and the arrangement under consideration can not be realized. So we see, that the choice of the boundary values from Table 1 is the only one, which leads to the unit baryon charge and finite energy.

\subsection{Numerical study}
To check whether the solution with boundary values considered in the previous section exists, calculate its energy and study the dependence of the mass of the baryon on the chiral condensate we rely on numerical analysis. Although the phase variables (\ref{phases}) were convenient to analyze the vacua of the model and construct the boundary asymptotics, they are not suitable for the numerical calculation. Thus we use the scalar fields introduced in (\ref{ansatz}) and solve their equations of motion, which follow from the action (\ref{action-scalars}). We choose the boundary conditions defined by the values of phases in Table 1. We summarize these conditions in Table 2.

\begin{table}[h!]
\begin{tabular}{|c|c|c|c|c|}
 \hline
Boundary & $z=0$ &  $r=0$ & $r \rar \infty$ & $z=z_m$  \\ \hline
$ \eta_1$ & $0$ & $0$ &  $0$ & $0$ \\ \hline
$ \eta_2$ & $-1$ & $-\cos(\theta(z))$ &  $-\cos(\theta(z))$ & $0$ \\ \hline
$ \xi_1$ & $0$ & $ \sin(\theta(z))$ & $ -\sin(\theta(z))$ & $-\sin(\omega(r))$ \\ \hline
$ \xi_2$ & $0$ & $0$ & $0$ & $-\cos(\omega(r))$ \\ \hline
$ \chi_1$ & $0$ & $-z^3 \sigma$ & $z^3 \sigma$ & $z^3 \sigma \sin(\omega(r))$ \\ \hline
$ \chi_2$ & $0$ & $0$ & $0$ & $z^3 \sigma \cos(\omega(r))$ \\ \hline
$A_r$ & $0$ & $0$ & $0$ & $-\p_r \omega(r)$ \\ \hline
$A_z$ & $0$ & $0$ & $0$ & $0$ \\ \hline
$V_r$ & $0$ & $0$ & $0$ & $0$ \\ \hline
$V_z$ & $0$ & $\p_z \theta(z)$ & $-\p_z \theta(z)$ & $0$ \\ \hline
\end{tabular}
\caption{The boundary values for the scalar fields, which describe the solution with unit baryon charge. $\theta$ and $\omega$ are described in text.}
\end{table}

We use the value for the $\chi$ (\ref{chi}) in the chiral limit. All boundary conditions are defined by the two functions $\theta(z)$ and $\omega(r)$. Their form is arbitrary apart of the boundary values
\begin{align}
\theta(0) &= 0, &  \theta(z_m) &= \frac{\pi}{2}, \\
\omega(0) &= -\frac{\pi}{2}, &  \omega(\infty) &= \frac{\pi}{2}.
\end{align}
Moreover, the regularity of the energy and the absence of sources require that the first and second derivatives of these functions are zero on the boundaries. Hence, the asymptotic values of the solution, which we look for, are characterized by the two step-like functions. We can introduce the parameters describing positions of ``steps'' and their widths. These four parameters would be the moduli of the instanton: its position on 2D plane and its diameters in two directions. However, in our model these parameters are not the moduli in the full sense, because the energy of the solution depends on them.

In our numerical study we construct the solution to equations of motion, which follow from (\ref{action-scalars}), with the boundary conditions from Table 2 for the particular choice of the position and diameters. Given the solution we calculate its energy. Then we repeat this procedure for the different choices of the parameters and try to find the solution with the minimal energy. Thus we perform the scan of the four dimensional parameter space in order to find the true stable solution. In the end of the day we find the solution with minimal energy for a number of values of $\sigma z_m^3$, which is the only dimensionless parameter of the model. The details of the numerical study can be found in the Appendix.

\begin{figure}[h!]
\includegraphics[width=0.7 \linewidth]{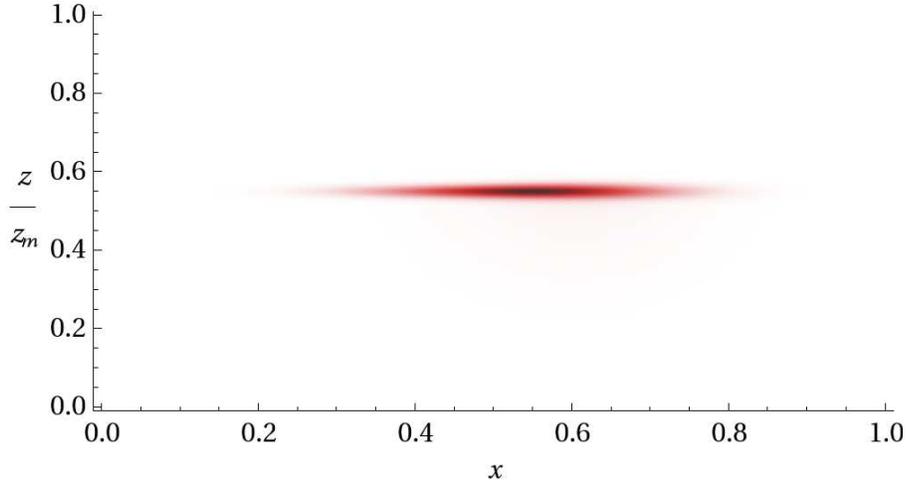}
\caption{\label{disk} The energy density of the solution with small $z$ radius and large $r$ radius. The $r$ coordinate is rescaled so that $r \rar \infty$ boundary is located at $x=1$.}
\end{figure}

A few interesting facts can be observed while searching for the optimal values of the moduli of the solution. First of all we find, that the solution tend to have a smallest possible radius in the $z$ direction and the largest possible in the $r$ direction. That means it looks more or less as a thin disk (see Fig. \ref{disk}). This form reminds us of the solution considered in our previous work \cite{gk}, where the domain wall like configuration was obtained. Thus our present solution preserves the analogy pointed out in \cite{gk} with the picture of the Skyrmion, catched on the domain wall \cite{Atiah-Manton}.

Even more interesting is the behavior of the optimal coordinates of the solution. While the $r$ coordinate is fixed by the balance of the large derivatives in the core of the instanton (at $r=0$) and large volume on the outskirts (at $r \rar \infty$) and doesn't show any dependence on the $\sigma$, the $z$ coordinate of the solution is tightly related to the value of the chiral condensate. We find, that for large values of $\sigma$ the optimal position of the instanton is located somewhere between IR and UV boundaries of the bulk space. This is in contrast with the usual behavior of the instanton solutions in $AdS$ \cite{Sakai-baryon, Wulzer}, which tend to fall on the IR boundary because due to the metric factor $\frac{1}{z}$ the energy is less for large values of $z$. In our treatment we find, that the chiral condensate provides a counter force, which drives the solution from IR boundary into the bulk. The larger $\sigma$ we take, the lower is the optimal position of the instanton. On the other hand, when 
we consider smaller $\sigma$, solution tends to stabilize at the larger $z$ until it finally touches the boundary. Further decreasing $\sigma$ gives no effect, because the solution can not get closer to the boundary as it would require diminishing its size and will lead to the growth of the kinetic part of the energy. This treatment shows, that the chiral condensate and the related bifundamental field $X$ play the important role of stabilizing the position of the instanton solution, which we consider in this work. This points to the fundamental relation between baryon and chiral physics in QCD.

\begin{figure}[h!]
\includegraphics[width=0.7 \linewidth]{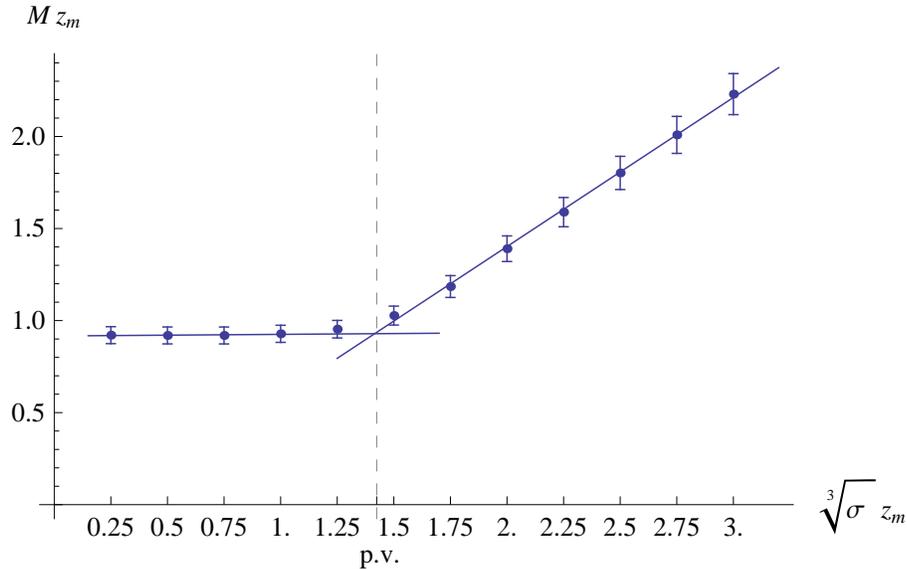}
\caption{\label{masses} The dependence of the baryon mass on the chiral condensate. The lines are constructed by the least squares fit. The fit is well inside the 5\% precision error bars of the energy values. The dashed line represents the physical values (p.v.) of the parameters, which provide the fit of the hard wall model to the real observables.}
\end{figure}

The fact that the chiral condensate governs the $z$ position of the solution and the energy of the solution that is the mass of the  baryon  suggests to study the dependence of the baryon mass on the chiral condensate in our model. On Fig. \ref{masses} we plot the energy (mass) of the optimal solution with respect to the cube root of $\sigma$, everything normalized to the position of the IR boundary $z_m$. Again we see  that for small $\sigma$ the energy does not change since the solution lyes just on the IR boundary, but for larger $\sigma$ the solution detaches from the boundary and the mass start to grow with the chiral condensate. The striking feature of this plot is the obviously linear dependence of the baryon mass on the energy scale of the condensate. This fits with the Ioffe's formula
with the additional constant term
\cite{Ioffe}
\begin{equation}
\label{result}
M_b = \mathrm{Max}\big[(a \la \bar{q} q \ra^{1/3} + b), \ c \big].
\end{equation}
We find  the approximate value  $a \approx 1.72$, $b \approx - 70 Mev$, $c \approx 300 Mev$, where the normalization (\ref{sigma}) is used. This formula underestimates the mass of the physical baryon by the factor of 3 and this effect may be attributed to the uncertainties in the parameter values in the ``hard-wall'' model or to the oversimplification of the bulk metric and the $N_f=2$. Nevertheless, we stress the linear dependence observed in (\ref{result}), which is in qualitative agreement with the (\ref{Form_Ioffe}), (\ref{linear}). It provides the very interesting resolution
of the longstanding puzzle concerning the origin of the Ioffe's formula.
We shell discuss the interpretation of this result more in the last section.

\section{Dyonic instanton in the brane picture \label{3}}

In this Section we shall make a few general comments concerning the dyonic instanton realization of the baryon
and the states with higher baryonic charges. First, remind the initial brane realization of the Lambert-Tong dyonic instanton \cite{dyonic2, hashimoto}. It is represented by the two parallel D4 branes displaced in some direction at the distance identified with the scalar vev. There are D0 branes localized at the D4 branes and F1 strings connecting
D0 branes. This configuration is unstable and decay into the tubular D2 brane (or equivalently D2-$\bar{D2}$ pair) which carries the flux due to dissolved D0 brane and  the Kaluza-Klein modes due to F1 strings.  The coordinate at the Coulomb branch in 5D Super Yang-Mills theory fixes
the asymptotic value of the scalar and the instanton size and mass. The key point of the solution is the effective 4D electric charge which is fixed by the boundary condition and prevents the solution from shrinking.

In the case of the flavor gauge group the theory is defined on the worldvolume of D8-$\bar{D8}$ pair wrapped around $S^4$ and extended along the radial coordinate of the cigar geometry \cite{ss}. The left and right D8 branes are connected in the chirally broken phase. What is the brane content of the flavor dyonic instanton? The conventional instanton-Skyrmion  is represented by D4 brane wrapped around
$S^4$ \cite{witten1, witten2} and the additional ingredient is the F1 string connecting the wrapped D4 branes. The asymptotic distance between the left and right flavor branes is fixed by the chiral condensate which is properly normalized vacuum expectation value of the bifundamental scalar \cite{berg} similar to the case of the color dyonic instanton. As in the color group case we expect the blow up of D4 branes into the tubular D6 brane with zero total D6 charge. Such blow up process is responsible for the Callan-Rubakov effect of the baryon decay induced by the monopole \cite{hashimoto}.

Could we make some additional qualitative remarks having in mind this brane realization? First note that upon the chiral restoration phase transition the D8 branes get disconnected since the cigar configuration is substituted by the cylinder. It means that the dyonic instanton solution becomes more similar with the color dyonic instanton.
The asymptotic value of the adjoint scalar is now the new parameter which fixes the baryon size upon the phase transition. Naively the F1 string stretched between two D8 branes amounts to their attraction and instability.
However this point deserves for the additional study.

Another point to be mentioned is the $\theta$-dependence. In the Witten-Sakai-Sugimoto like picture the $\theta$-term is introduced as the holonomy of the one-form RR field along the compact coordinate of the cigar at infinity
\beq
\theta =\oint _{\infty} C_1
\eeq
This results into the interesting possibility for $\theta$ dependence of the baryon. For the conventional Skyrmion there is no $\theta$ dependence since the D4 branes are localized at the fixed point in the angular coordinate  of the cigar.
However since the D4 branes are blown up into
the D6 brane we could ask if the $C_1$ field interacts with the gauge fields at the D6 worldvolume and enters the
D6 worldvolume action. Since from the four-dimensional viewpoint the D6 brane looks like
a magnetic string we could speak on the worldsheet theory for this string.

Let us consider the
CS terms in the  6+1 dimensional  worldvolume theory on the D6 brane
\beq
L_{CS}= \int _{D6} C_3 \wedge F \wedge F +    \int _{D6} C_1\wedge F \wedge F \wedge F
\eeq
The first term amounts to the CS term on the D6 action with coefficient $N_c$. The second term
is more interesting and generates the $\theta$ -term at the worldsheet of the magnetic string
with the baryonic charge similar to the picture developed in \cite{gz, gz2}. Hence at least for the
state with B=2 with the expected toroidal structure we expect the nontrivial sensitivity to the
bulk $\theta$-term.

\section{Discussion \label{4}}

The key finding of our paper is that the infinite tower of mesons packed into the holographic gauge
field with the flavor group as well as the chiral condensate are equally important for the
determination of the baryon mass. The role of the chiral condensate is more
complicated then naively Ioffe's formula implies. It is not possible to claim that
the mass of the baryon is completely fixed by the quark condensate, because it provides only
one of two competing contributions.

Going back to the 80-ties recall that the derivation of the
baryon mass \cite{Ioffe} in the sum rules approach is based on the consideration
of the correlator of the three-quark baryonic current. The leading contribution turns out
to be obtained from the quark condensate multiplied by the loop involving the two-quark state
whose imaginary part involves the infinite tower of resonances.
This realization has some similarity with our picture. Indeed we have found that
both chiral condensate and the tower of mesons are equally important. It would be very interesting
to support these naive arguments by the holographic calculation of the two-point correlator
of the baryonic currents.

The behavior which we observe on Fig. \ref{masses} shows  that the mass of the baryon is defined by the competition of two different mechanisms: one related to the conformal symmetry breaking, which is described in our model by the IR hard wall and another -- the chiral condensate. When chiral condensate is small, the mass of the baryon is defined by the warp factor of the metric pulling the solution to the IR hard wall and the position $z_m$ of this wall. However when the condensate becomes large the position of the solution is governed by the interaction with the $X$ field and the value of $\sigma$. In this context we should pay the attention to the amazing fact, that the point on Fig.\ref{masses}, where the lines describing different regimes intersect, namely the point, where two mechanisms are balanced, almost exactly coincide with the values of parameters, which provide the fit of the hard wall model to the real physical observables \cite{hard-wall, Krik1}
\begin{equation}
(\sigma)^{1/3} z_m = \frac{460 Mev}{323 Mev} \approx 1.42.
\end{equation}
This fact could be considered as some evidence that  the scale of chiral symmetry breaking and the scale of conformal symmetry breaking are related in such a way that provides the minimum baryon mass.

At a moment we could only speculate about the precise mechanism responsible for this competition. Let us remind that
the baryon mass  can be obtained from the conformal anomaly determined by the
trace of the energy-stress tensor $T_{\mu,\mu}$
\beq
M_B\propto \la B|T_{\mu\mu}|B \ra
\eeq
It is well-known that the conformal anomaly involves the gluon and quark contributions
\beq
T_{\mu\mu}= \beta TrG^2 + m \bar{q}q
\eeq
where $\beta$ - is the beta-function of the gauge theory.
Widely appreciated viewpoint is that in the chiral limit we can neglect
the second term and the baryon mass is determined purely by the gluon component
of the baryon or the gluon condensate. However our analysis implies that such
naive picture is oversimplified. It seems that the gluon condensate inside the baryon is
nontrivial function of the chiral condensate and just minimization of the energy
with respect to the chiral condensate provides the baryon mass. The interplay between
the gluon and quark condensates is known at  large quark mass and is responsible
for the decoupling of the heavy flavor. The decoupling is valid in the
holographic models \cite{decoup} indeed however it is not clear how it could work
in our case in the dyonic instanton background. Probably the recent 
discussion in \cite{rho2} is also relevant for this problem.

There are a lot of questions which have to be elaborated within our approach. First of all
it would be very interesting to analyze in details the solutions with higher baryonic charges.
We expect to reproduce the toroidal structure of the $B=2$ solution known for a while
in the low-energy QCD. It fits with the toroidal structure of the dyonic instanton
with instanton number $Q=2$. It would be very interesting to investigate the
amount of the proton spin carried by the quarks. The dyonic instanton is known to
have very subtle contributions to the angular momentum. Another interesting issue concerns
the physical characteristics of the baryon like  formfactors or higher tensor charges.
This would demand to add the tensor deformation to the holographic model discussed
in \cite{karch, tensor, harvey}. 
Another interesting possibility concerns the application of our analysis to the 
Skyrmions in the solid state framework using the $AdS_4$ like geometry.
Finally mention the possibility to analyze the scenario
for the baryon driven chiral phase transition  \cite{kogan} in  our framework.
We plan to discuss these issues elsewhere.

\acknowledgments
 We are grateful to Dmitry Kharzeev and nuclear physics group of the Stony Brook University for the useful discussions, comments and  the hospitality. A.K. thanks Alexander Rokhmanenkov for valuable advices concerning the numerical calculus. The work of A.K. and P.N.K. is partially supported by the Ministry of Education and Science of the Russian Federation under contract 14.740.11.0347 and the Dynasty Foundation. A.G.
 thanks FTPI at University of Minnesota where the part of the work has been done for the hospitality and support.
The work is partially supported by RFBR grant no.12-02-00284 and PICS- 12-02-91052.
\appendix
\section{Numerical calculation}
We use numerics to obtain the solution to the equations of motion, which follow from (\ref{action-scalars}) with boundary conditions specified in Table 2. First of all we rescale the domain where the solution is calculated by defining the dimensionless coordinates
\begin{align*}
\tilde{z} &= \frac{z}{z_m} \\
x &= \frac{2}{\pi} \arctan \Big(\frac{\pi}{2} \frac{r}{c} \Big).
\end{align*}
This reduces the domain to the unit square and allows us to include the boundary $r \rar \infty$. Similarly we introduce the dimensionless parameter $\tilde{\sigma} = \sigma z_m^3$. Note, that the dimensional constant $c$ defines the region in $r$ which is best resolved. Later we will study the solution located at $x=0.5$ and $c$ will actually define its position in $r$, namely the radius of the baryon.

After rescaling all the coordinates we observe that the equations of motion we are about to solve are singular at three of four boundaries of the integration region, namely $\tilde{z}=0, x=0$ and $x=1$. Thus we are required to use extreme caution when dealing with these boundaries.

In the gauge (\ref{alpha}) the equations of motion are elliptic and this allows us to use the relaxation procedure to define the solution by putting the equations on the grid and substituting the derivatives with the finite differences. We start with the initial guess, which is compatible with the boundary conditions in Table 2. This is
\begin{align*}
\eta_1^{(0)} (\tz, x) &= 0, & \eta_2^{(0)}(\tz,x) &= -\cos(\theta(\tz)), \\
\xi_1^{(0)}  (\tz, x) &= \sin(\theta(\tz)) \sin(\omega(x)), & \xi_2^{(0)}(\tz,x) &= \sin(\theta(z)) \cos(\omega(x)), \\
\chi_1^{(0)}  (\tz, x) &= \tz^3 \ts \sin(\omega(x)), & \chi_2^{(0)}(\tz,x) &=  \tz^3 \ts \cos(\omega(x)), \\
A_r^{(0)}  (\tz, x) &= -\frac{\p_r}{\p_x} \ \p_x \omega(x) \sin(\theta(\tz))^2, & A_z^{(0)}(\tz,x) &=  0, \\
V_r^{(0)}  (\tz, x) &=0,  & V_z^{(0)}(\tz,x) &= - \sin(\omega(x))  \p_{\tz}  \theta(\tz).
\end{align*}

As was already pointed out in the main text, the functions $\omega$ and $\theta$ are step-like. They assume definite values at the boundaries and have vanishing derivatives there. Firstly we assumed them to behave as an error function (the integral of the Gauss), because it has well defined center and the width. For $\theta(z)$ this turns out to be a good choice, because we can control well the position and take quite a small radius of the solution -- we are able to take it approximately $0.05$ (see Fig. \ref{disk}). For the $\omega(x)$ it turns out, that the energy falls with enlarging width of the step until the width is of order 1. That tells us, that the exponential jump in the step is too fast. Moreover, we do not need to control the position of the step along $x$, because we have another parameter $c$, which does the same job in the variable $r$. So instead of exponential error function we use the polynomial step, which is located at $x=0.5$, and control the position of the solution by $c$.

Instead of discretizing expressions for initial guess on a grid and using them as the initial values for the relaxing functions we substitute the initial functions  directly to the equations before introducing the grid and solve the equations for the deviations from this initial guess. This allows us to get rid of possible problems with discretizing derivatives on the singular boundaries. Hence we expand the functions as
\begin{align*}
\eta_1  (\tz, x) & = \eta_1^{(0)} (\tz, x) + \frac{1}{x^{\frac{3}{2}}} \tilde{\eta}_1 (\tz, x), &
\eta_2  (\tz, x) & = \eta_2^{(0)} (\tz, x) + \frac{1}{x^{\frac{3}{2}}} \tilde{\eta}_2 (\tz, x), \\
\xi_1  (\tz, x) & = \xi_1^{(0)} (\tz, x) + \frac{1}{x^{\frac{3}{2}}} \tilde{\xi}_1 (\tz, x), &
\xi_2  (\tz, x) & = \xi_2^{(0)} (\tz, x) + \frac{1}{x^{\frac{3}{2}}} \tilde{\xi}_2 (\tz, x), \\
\chi_1  (\tz, x) & = \chi_1^{(0)} (\tz, x) + \tz \tilde{\chi}_1 (\tz, x), &
\chi_2  (\tz, x) & = \chi_2^{(0)} (\tz, x) + \tz \tilde{\chi}_2 (\tz, x), \\
A_r  (\tz, x) & = A_r^{(0)} (\tz, x) + \frac{1}{x} \tilde{A}_r (\tz, x), &
A_z  (\tz, x) & = A_z^{(0)} (\tz, x) +  \tilde{A}_z (\tz, x), \\
V_r  (\tz, x) & = V_r^{(0)} (\tz, x) + \frac{1}{x} \tilde{V}_r (\tz, x), &
V_z  (\tz, x) & = V_z^{(0)} (\tz, x) + \tilde{V}_z (\tz, x). 
\end{align*}
(The rescalings are introduced to handle the singular boundaries.) After this substitution the problem is reduced to finding the solutions for the fields with tildes subject to the zero boundary conditions. We introduce the grid and consequently solve the discretized equations in each vertex to obtain the new values of the relaxing functions. We do not solve the equations on the boundary, as they are automatically satisfied by the boundary conditions. The procedure is repeated until the new values of the relaxing functions deviate from the previous ones less then in 5\%. The resulting discrete functions are interpolated in order to compute the energy of the baryon via the integral (\ref{action-scalars}).

\end{document}